\documentclass[fleqn,usenatbib]{mnras}
\usepackage[T1]{fontenc}
\usepackage{graphicx}
\usepackage{amsmath}
\usepackage{amssymb}
\usepackage{newtxtext,newtxmath}

\title[Separate ICs in Ly$\alpha$ Simulations]{Effect of Separate Initial Conditions on the Lyman-$\alpha$ Forest in Simulations}
\author[M.A. Fernandez et al.]
{
M.A. Fernandez,$^{1}$\thanks{mfern027@ucr.edu}
Simeon Bird,$^{1}$\thanks{simeon.bird@ucr.edu}
and Phoebe Upton Sanderbeck$^{1}$\thanks{phoebeu@ucr.edu}
\\
$^{1}$Department of Physics and Astronomy, University of California Riverside, 900 University Ave, Riverside, CA 92521
}
\date{Accepted XXX. Received YYY; in original form ZZZ}
\pubyear{2020}

\begin{document}
\label{firstpage}
\pagerange{\pageref{firstpage}--\pageref{lastpage}}
\maketitle
\begin{abstract}
Using a set of high resolution simulations, we quantify the effect of species specific initial transfer functions on probes of the IGM via the Lyman-$\alpha$ forest.
We focus on redshifts $2-6$, after H~{\sc i} reionization.
We explore the effect of these initial conditions on measures of the thermal state of the low density IGM: the curvature, Doppler width cutoff, and Doppler width distribution.
We also examine the matter and flux power spectrum, and potential consequences for constraints on warm dark matter models.
We find that the curvature statistic is at most affected at the $\approx2\%$ level at $z=6$.
The Doppler width cutoff parameters are affected by $\approx5\%$ for the intercept, and $\approx8\%$ for the fit slope, though this is subdominant to sample variation.
The Doppler width distribution shows a $\approx30\%$ effect at $z=3$, however the distribution is not fully converged with simulation box size and resolution.
The flux power spectrum is at most affected by $\approx5\%$ at high redshift and small scales.
We discuss numerical convergence with simulation parameters.
\end{abstract}


\section{Introduction}\label{sec:intro}

The intergalactic medium (IGM) occupies the space between galaxies and galaxy clusters, and houses the majority of baryonic matter in the universe.
The major phase changes in the history of the IGM are fairly well understood, with recombination ($z\sim1100$) leading to the formation of a highly neutral IGM, and H~{\sc i} ($z\sim 5.5-8$) \citep{2006AJ....132..117F, robertson10,wmap, 2018arXiv180706209P, 2019ApJ...872..101B} and He~{\sc ii} ($z\sim3$) \citep{madau99,miraldaescude00, wyithe03, furlanetto08, shull10, 2016ApJ...825..144W} reionization events leading to the current, highly ionized IGM \citep[for a review on the IGM, see][]{2016ARA&A..54..313M}.
The sources of the ionizing photons are thought to be stars in galaxies \citep{2016ApJ...831..176B}, and quasars \citep{madau99,2009ApJ...694..842M, haardt12} for H~{\sc i} and He~{\sc ii} reionization, respectively.

During reionization, ionizing photons heat the IGM by tens of thousands of degrees.
This heating, combined with cooling from adiabatic expansion and atomic processes, are the primary processes that influence the thermal state of the low density ($1-100$ times the cosmic mean density) IGM \citep{1994MNRAS.266..343M, 1997MNRAS.292...27H, schaye00, hui03, 2016MNRAS.460.1885U, 2019ApJ...874..154D}.
The thermal energy of the IGM smooths and extends the distribution of the gas, which in turn affects structure formation.
After each reionization event, the low density IGM cools asymptotically towards an equilibrium temperature \citep{1997MNRAS.292...27H, 2016MNRAS.456...47M}.
During this time the ionization state is well understood, as the neutral fraction is set by the equilibrium between photoionizations and recombinations.
All of this makes the IGM, and especially the low density IGM, a valuable probe of the post-reionization universe ($z<6$) and the scales probed make it useful for both astrophysics and cosmology.

Conveniently, there are numerous observations probing intergalactic gas at $2<z<6$.
Generally, these are observations of the Lyman-$\alpha$ forest, the series of absorption features blueward of the rest-wavelength Lyman-$\alpha$ emission observed in quasar spectra \citep{1965ApJ...142.1633G}.
A single forest spectrum is a one-dimensional map of the gaseous structure along that line of sight, making it a useful probe of structure formation.
Knowledge of the large scale structure, either through the flux power spectrum or the inferred matter power spectrum, constrains warm dark matter (WDM) models \citep{2005PhRvD..71f3534V, 2019ApJ...872...13W}.
In addition to probing structure formation, the Lyman-$\alpha$ forest can be used to measure the thermal state of the IGM, leading to a set of measurements describing the thermal history of the IGM.
Using the thermal and ionization history of the IGM, one can test models of the makeup and evolution of the ionizing background, and thus infer properties of the ionizing sources and sinks over time \citep{2019ApJ...872..101B}.

There are several ways in which Lyman-$\alpha$ forest spectra are processed to constrain cosmological models and the thermal state of intergalactic gas.
Cosmological contexts generally make use of the flux power spectrum from a sample of Lyman-$\alpha$ forest spectra \citep{2001ApJ...557..519Z, 2013A&A...559A..85P, 2016MNRAS.463.2335N, 2019ApJ...872..101B}.
The flux power is the Fourier transform of the flux over-density, $\delta_F = F/\langle F \rangle - 1$.
The flux power spectrum is sensitive to cosmological parameters on large scales ($k < 0.02$ s/km for velocity wavenumber $k$), and constrains small scale smoothing at higher $k$ \citep{2015ApJ...812...30K}.
For example, smoothing is enhanced in WDM models, leading to a reduction in power above some critical value of $k$, (dependent on the mass of the WDM particle).
This makes the flux power spectrum a robust tool for constraining WDM models \citep{2019ApJ...872...13W}.

The spectral statistics used in determining the thermal state of the IGM are more varied.
Common methods include statistics which encapsulate an entire forest spectrum \citep{theunszaroubi, theuns02, zaldar02, lidz10, 2011MNRAS.410.1096B, boera14}, as well as analyses which make use of absorption features from spectra decomposed via Voigt profile fitting \citep{1999MNRAS.310...57S, 2000ApJ...534...41R, schaye00, 2001ApJ...562...52M, 2014MNRAS.438.2499B, 2018ApJ...865...42H}.
The small scale flux power spectrum and the distribution of flux are also used to constrain the IGM thermal state \citep{2001ApJ...557..519Z,2020arXiv200900016G}.

The Lyman-$\alpha$ forest probes scales on which non-linear structure growth is important, and so cosmological hydrodynamic simulations of the IGM are necessary to build a map between model parameters and observations.
These simulations require two components: collisionless cold dark matter modelled using N-body techniques, and collisional baryons which include pressure forces.
One common simplification is that, although baryons are evolved hydrodynamically, the initial conditions for both species are identical, using the transfer function for the total matter fluid \citep{2019ApJ...877...85E}.

Before recombination, baryons couple to radiation, suppressing their clustering on sub-horizon scales and reducing clustering relative to the dark matter.
After recombination, baryons fall into the potential well of the cold dark matter and so the linear transfer functions differ by $<1$\% at $z=0$.
The effect is larger at higher redshifts, $z=2-5$, where the Lyman-$\alpha$ forest is a sensitive probe of the gas \citep{2005MNRAS.362.1047N}.
\cite{2020arXiv200200015B} showed that separate transfer functions can affect the one-dimensional Lyman-$\alpha$ forest flux power spectrum by $5-10\%$ on scales $0.001 - 0.01$ s/km in the redshift range $z=2-4$.

The aim of this work is to determine whether species specific initial transfer functions have an appreciable effect on probes of the Lyman-$\alpha$ forest.
We use the simulation technique developed in \citet{2020arXiv200200015B}, which reproduces the theoretical offset between the dark matter and baryon power \citep{2013MNRAS.434.1756A}, to model separate initial transfer functions.
Recently, \citet{2020arXiv200809123R} \citep[see also][]{2020arXiv200809124H, 2020arXiv200809588M} resolved this discrepancy by perturbing the particle masses, in agreement with the results from \citet{2020arXiv200200015B}.
We will examine the effect of these initial conditions on measures of the thermal state of the IGM; the curvature \citep{2011MNRAS.410.1096B}, Doppler width cutoff \citep{1999MNRAS.310...57S}, and Doppler width distribution \citep{2020arXiv200900016G}.
We also examine the effect on the matter and flux power spectrum, which could have consequences for warm dark matter models \citep{2000ApJ...543L.103N}.
The simulations we use are higher resolution than in \citet{2020arXiv200200015B}, allowing us to better probe smaller scales.

In Section \ref{sec:simulations} we outline the simulations and artificial spectra used throughout.
In Section \ref{sec:methods_results} we discuss the methods used to calculate each measure of the IGM, as well as the results of those calculations.
Measures of the thermal history of the IGM, including the curvature, and the Doppler width cutoff and distribution, are covered in sections \ref{sec:curvature} \& \ref{sec:doppler}, respectively.
The WDM relevant measures are examined in Sections \ref{sec:fluxps} (flux power spectrum) and Section \ref{sec:matterps} (matter power spectrum).
In Section \ref{sec:conclusions} we summarize and conclude.
We include Appendix \ref{appendix:convergence}, which discusses numerical convergence with box size,  resolution, and number of artificial spectra used.

We assume throughout a flat $\Lambda$CDM cosmology with $\Omega_0=\Omega_b+\Omega_{CDM}=0.288, \Omega_b=0.0472, h=0.7, n_s=0.971,$ and $\sigma_8=0.84$ \citep[consistent with 9-year WMAP results][]{2013ApJS..208...19H}.

\section{Simulations}\label{sec:simulations}

Our set of hydrodynamical simulations were performed using the N-body and smoothed particle hydrodynamics (SPH) code MP-Gadget\footnotemark, described in \citet{2018MNRAS.481.1486B, 2019JCAP...02..050B}.\footnotetext{\url{https://github.com/sbird/MP-Gadget3}}
MP-Gadget is a fork of Gadget-3, itself the descendent of Gadget-2 \citep{2005MNRAS.364.1105S}.
Initial conditions are generated with MP-GenIC, the initial conditions generator packaged with MP-Gadget.
The initial power spectrum, and transfer functions are generated with the Boltzmann code CLASS \citep{2011arXiv1104.2932L}.

Simulations using offset grids for both particle species (which is common in the literature) often introduce a spurious growing mode to the CDM-baryon difference.
This can be avoided by using a glass to initialize the baryons \citep{2003MNRAS.344..481Y, 2020arXiv200200015B}, or by an appropriate perturbation of the particle masses \citep{2020arXiv200809124H}.
Two sets of simulations are used throughout this work, with initial conditions set using the baryon-glass method.
Both sets of simulations use a glass to initialize the baryons and a grid to initialize the CDM.
A glass procedure, with 14 time-steps, is then applied to the combined distribution to minimize CDM-baryon overlap, avoiding chance overdensities set by the initialization.
The two sets of simulations then differ, with the first set using a single transfer function for both species, and the second set using separate, species specific transfer functions.
Scale-dependent perturbations are included via first-order Lagrangian perturbation theory (during final preparation of this manuscript, \citet{2020arXiv200809124H, 2020arXiv200809123R} proposed an alternative method based on second-order perturbation theory, which gives similar results).
The phases of the Fourier modes are identical, leading to the same realization of cosmic structure on scales larger than the particle grid.

Gas is assumed to be in ionization equilibrium with a uniform ultraviolet background using the model of \citet{2009ApJ...703.1416F}\footnotemark.
\footnotetext{Specifically the 2011 update, \url{https://galaxies.northwestern.edu/uvb-fg09/}}
\citet{2020MNRAS.493.1614F} recently updated their UV background model and showed that simulations using uniform UV backgrounds do not accurately model the timing and photoheating associated with reionization.
In our simulations reionization has completed by $z=6$ (the average neutral hydrogen fraction in low density regions of our simulations is less than $1\%$).
Our results are generated in the redshift range $2 < z < 6$, after hydrogen reionization.
We do not implement He~{\sc ii} reionization because the scale of our simulation box size is smaller than a typical He~{\sc ii} bubble \citep{2020MNRAS.496.4372U}, leading to an effectively instantaneous reionization.

Star formation is implemented using the standard approach for Lyman-$\alpha$ forest analyses.
Gas particles in the simulations are turned into stars using a simple density-based method; when they reach an overdensity $\rho/\langle\rho\rangle > 1000$, but remain at a temperature $T < 10^5$, they are turned into stars \citep{2004MNRAS.354..684V}.
Our simulations do not include black hole or supernovae feedback.

The set of high-resolution simulations include our Main simulations, a simulation with lower gas mass resolution, and a simulation with a smaller box length of $10$ Mpc h$^{-1}$.
The latter two are used to check box size and gas mass resolution convergence.
We discuss convergence for each result in the relevant results section, as well as in Appendix~\ref{appendix:convergence}.
All simulations start at $z=99$ and have periodic boundaries.
Box volume, particle number, and gas particle mass resolution are reported in Table~\ref{table:simulations}.
The gas particle mass resolution is set so that the higher redshift Lyman-$\alpha$ forest is resolved \citep{2009MNRAS.398L..26B}.

\begin{table}
	\centering
	\caption{Simulations}
	\label{table:simulations}
	\begin{tabular}{lccc}
		\hline
		Simulation & Box Volume & N & M$_{\text{gas}}$ (M$_{\odot}$)\\
		\hline
		Main & $(20$ Mpc h$^{-1})^3$ & $2\times1024^3$ & $9.8\times10^4$\\
		Low Res & $(20$ Mpc h$^{-1})^3$ & $2\times768^3$ & $2.3\times10^5$\\
		Small Box & $(10$ Mpc h$^{-1})^3$ & $2\times512^3$ & $9.8\times10^4$\\
		\hline
	\end{tabular}
\end{table}

Lyman-$\alpha$ absorption spectra are generated by sending random skewers through the simulation box using Fake Spectra Flux Extractor \citet{2017ascl.soft10012B}\footnotemark, described in \citet{2015MNRAS.447.1834B}.
\footnotetext{\url{https://github.com/sbird/fake_spectra}}
Our analysis uses $5,000$ randomly placed skewers, which are generated for each snapshot, leading to a large set of $1$ km s$^{-1}$ pixel width neutral hydrogen absorption spectra for redshifts in the range $2 < z < 6$.
We discuss convergence of our results with number of sight lines in Appendix~\ref{appendix:convergence}.

\section{Methods \& Results}\label{sec:methods_results}

In this section we examine the effect using species-specific initial conditions has on two commonly studied properties of the IGM, both of which use Lyman-$\alpha$ forest spectra.
The first is the temperature-density relation of the low density IGM, which is generally parameterized as
\begin{equation}
    T(\Delta) = T_0\Delta^{\gamma-1},
\end{equation}
where $\Delta$ is the matter overdensity, $T_0$ is the temperature at mean density ($\Delta = 1$), and $\gamma-1$ is the power-law index \citep{1997MNRAS.292...27H, 2016MNRAS.456...47M}.
Throughout we focus on redshifts after H~{\sc i} reionization ($\leq 6$), where adiabatic cooling and photoheating dominate the thermal state.
This is the regime where the temperature-density relation parameterized above is generally valid (though it is not best described with a single temperature-density relation during He~{\sc ii} reionization) \citep{2008ApJ...689L..81T, 2009ApJ...701...94F,2020MNRAS.496.4372U}.
We focus on three measures which probe the temperature-density relation of the IGM: the curvature (\ref{sec:curvature}), the Doppler width cutoff (\ref{sec:b-parameter}), and the Doppler width distribution (\ref{sec:bpdf}).

The second property is the matter power spectrum of the IGM, which can constrain dark matter models, especially warm dark matter through its effect on structure formation.
The matter power spectrum of the dim and diffuse IGM is not directly accessible.
However, the flux power spectrum is a good proxy and allows constraints to be placed on the thermal free-streaming of dark matter and thus a potential WDM particle mass.
We examine the effect species specific initial conditions have on both the Lyman-$\alpha$ flux power spectrum (Section \ref{sec:fluxps}) and matter power spectrum (Section \ref{sec:matterps}).

\subsection{Curvature}\label{sec:curvature}

The curvature statistic introduced in \citet{2011MNRAS.410.1096B} has an approximately one-to-one relationship with the temperature of the IGM at an optimal overdensity.
The temperature at the mean density can then be inferred using a temperature-density relationship slope calibrated from simulations.
The curvature is essentially the second derivative, or curvature, of the flux.
Specifically, it is given by $\kappa \equiv F^{\prime\prime}/(1+{F^{\prime}}^2)^{3/2}$ and traces the ionized fraction of hydrogen.
Higher temperature gas will show more thermal broadening in the absorption features of the spectra, while lower temperature gas will retain more small-scale spectral features.
Because the curvature summarizes the entire spectrum it does not require decomposing spectra into individual absorbers, making it useful up to higher redshifts than the Doppler width methods (Section \ref{sec:doppler}).

\begin{figure}
    \centering
	\includegraphics[width=\columnwidth]{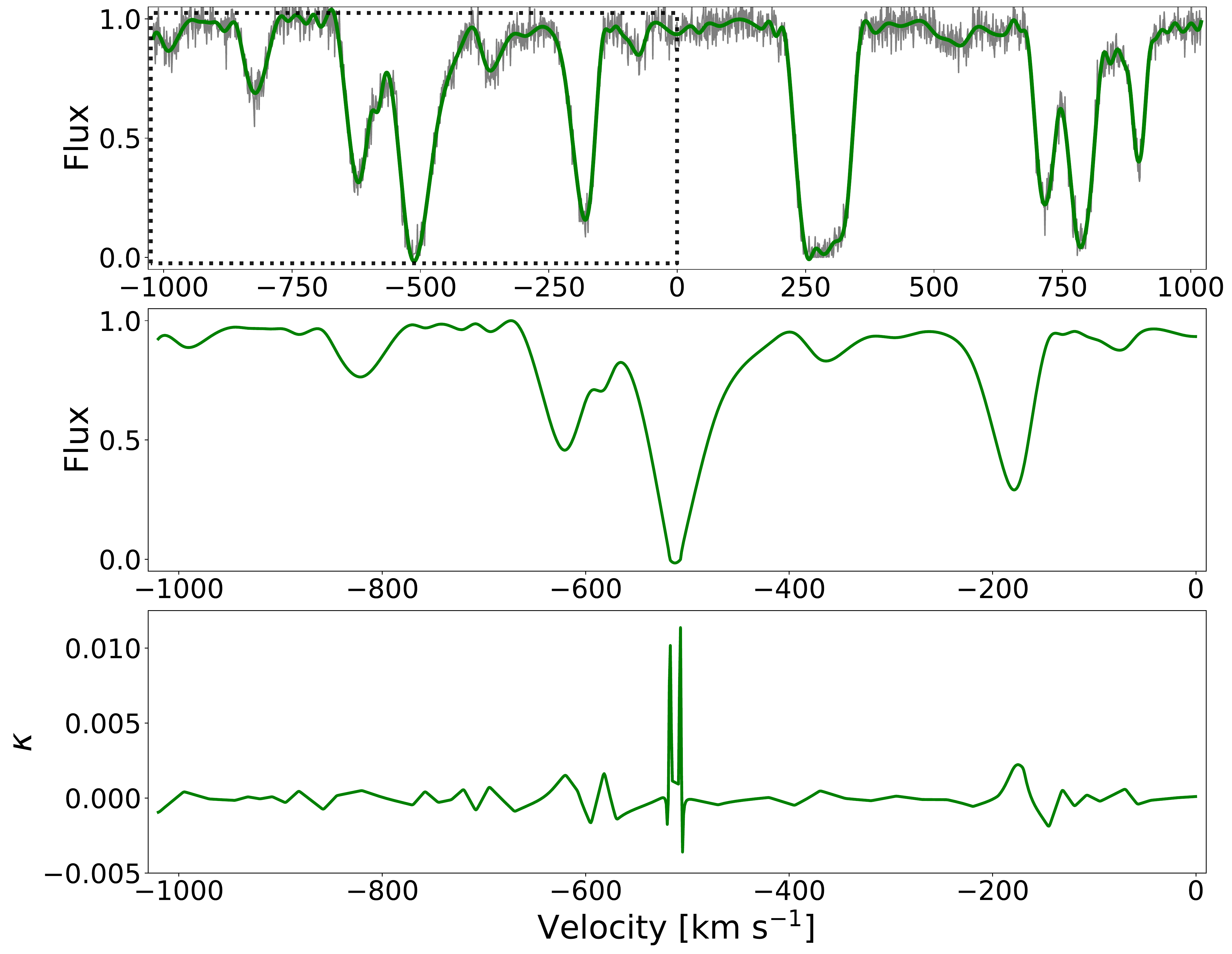}
    \caption{Example of the curvature measurement.
    Top: spectrum with noise (black) and the smoothed cubic spline fit to it (green).
    Middle: The region enclosed in the dotted box in the top panel, renormalized into $10$ Mpc h$^{-1}$ sections and rescaled such that the set of all these sections have the same mean flux across simulations.
    Bottom: the curvature of the middle panel.
    The single value reported for such a section is the mean absolute curvature value in regions where the renormalized and rescaled flux lies between $0.1 < F < 0.9$.}
    \label{fig:curvature_spectrum}
\end{figure}

The simulated spectra are processed, and the curvature calculated following the general procedure in \citet{2011MNRAS.410.1096B}:
\begin{enumerate}
    \item Gaussian noise is added to the spectra such that the $S/N \sim 20$, then a cubic b-spline is fit to the flux iteratively.
    The initial break point spacing between the piecewise b-spline is set at $50$ km s$^{-1}$ and additional points are added to improve the fit until either a minimum resolution is reached ($10$ km s$^{-1}$) or the fit converges (the $\chi^2$ value between spline and spectra changes by less than $3$ between break point additions).
    The resulting spline, an example of which can be seen in the top panel of Figure~\ref{fig:curvature_spectrum}, is used in place of the spectrum for the rest of the analysis.
    \item The spline is then renormalized by breaking it into $10$ Mpc/h sections and dividing by the maximum value in that section.
    This normalizes the measure and avoids uncertainties due to continuum finding.
    \item Each of these sections is then rescaled such that the mean flux of the entire set of sections is consistent with the model from \citet{2007MNRAS.382.1657K}, given by an effective optical depth, $\tau^{\text{eff}} = 0.0023(z+1)^{3.65}$.
\end{enumerate}

The result of steps (ii) and (iii) are shown in the middle panel of Figure~\ref{fig:curvature_spectrum}.
Note that the values used in the processing outlined above (e.g. the S/N, $\chi^2$ convergence value, etc.) are chosen either to agree with \citet{2011MNRAS.410.1096B}, to be reasonable in regards to observation, or simply to fit the artificial spectra well.

\begin{figure}
    \centering
	\includegraphics[width=\columnwidth]{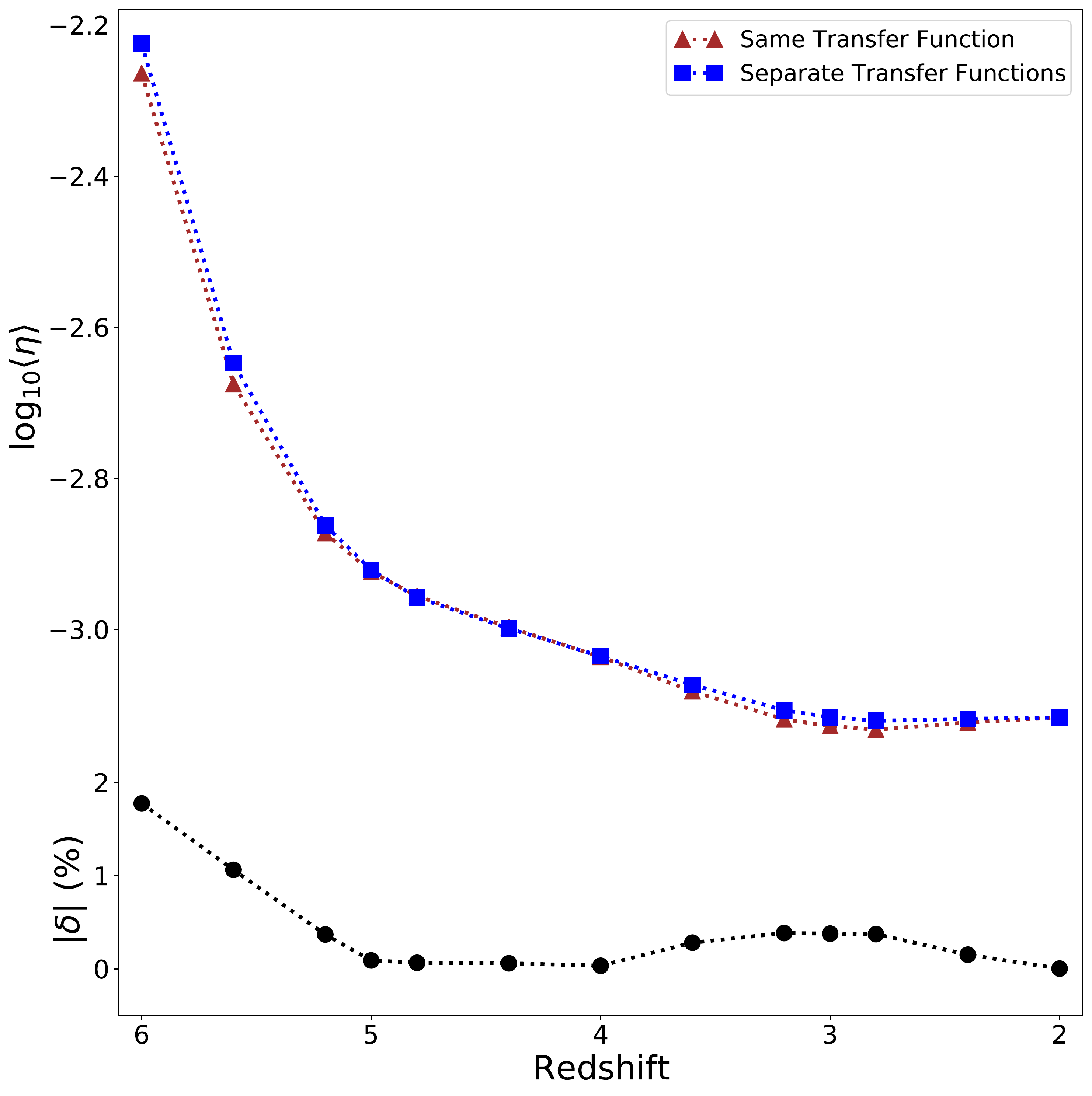}
    \caption{Results for the average curvature versus redshift.
    The curvature is robust up to higher redshift, hence the inclusion of redshifts up to $z=6$.
    The agreement is extremely good, with a maximum difference of $<2\%$ at redshift $z=6$.}
    \label{fig:curvature_results}
\end{figure}

The curvature is then calculated, using only flux in the range $0.1 < F < 0.9$.
The bottom panel of Figure~\ref{fig:curvature_spectrum} shows an example of the curvature, before restricting the flux range.
For each section the mean absolute curvature is returned, $\eta = \langle|\kappa|\rangle$, and the average of $\eta$ for each redshift is shown in Figure~\ref{fig:curvature_results} (top), along with the fractional difference between the two (e.g. $|x_1/x_2 - 1|$) in percentage (bottom).
The squares (blue) show the results for the simulation which uses separate initial transfer functions, and the triangles (brown) show the simulation which uses the same transfer function.
The difference between the two is remarkably small, peaking at $<2\%$ at $z=6$.

For convergence testing, noise is not added to the spectra, though the spectra are still renormalized and rescaled.
Convergence of the curvature is discussed in Appendix~\ref{appendix:convergence}, and shown in Figure~\ref{fig:curvature_convergence} for simulation parameters, and in Figure~\ref{fig:nlos_convergence_diff} (top panel) for convergence with number of sight lines used.

\subsection{Doppler Width Methods}\label{sec:doppler}

\subsubsection{Doppler Width Cutoff}\label{sec:b-parameter}

Another method used to determine the thermal state of the IGM, first introduced in \citet{1999MNRAS.310...57S}, is fitting the lower cutoff in the Doppler width ($b$) of spectral features as a function of their neutral hydrogen column density ($N_{\text{HI}}$).
The Doppler width characterizes the width of an absorption profile due to broadening from the motion of the particles constituting the absorber.
Theoretically, an absorber has a minimum Doppler width, i.e. due entirely to thermal broadening with no additional effects such as broadening from a velocity gradient.
The minimum will depend on the temperature of the absorber, $b_{\text{therm}} = \sqrt{2k_bT/m}$, where $k_b$ is the Boltzmann constant, $T$ the temperature, and $m$ the proton mass.
The temperature depends on the density, with higher density clouds having a higher temperature and thus a broader spectral profile.
As the true density of an absorber is not observable, the column density of the absorber and its Doppler width are used as proxies.
We then have a relation between the minimum Doppler width and the column density, where a higher column density absorber has a higher temperature, and thus a larger minimum Doppler width.
A lower cutoff, in the form of a power law, can then be fit to a set of Doppler width, column density measurements.
Using simulations and observations of the $N_{\text{HI}}-b$ cutoff parameters, the temperature-density relation is calibrated and the physical density inferred from the column densities \citep{2001ApJ...559..507S}.

This method requires decomposing spectra into features, with widths and amplitudes corresponding to the Doppler width and column density of the associated absorbers.
We use Fake Spectra Flux Extractor \citep{2017ascl.soft10012B} to decompose our artificial spectra, which are the optical depths ($\tau$) along lines of sight through the simulations, into individual features by fitting the flux ($F = e^{-\tau}$) using Voigt profiles.
The Voigt profile width is the Doppler width, $b$, and the Voigt profile normalized amplitude is the neutral hydrogen column density, $N_{\text{HI}}$.

\begin{figure}
    \centering
	\includegraphics[width=\columnwidth]{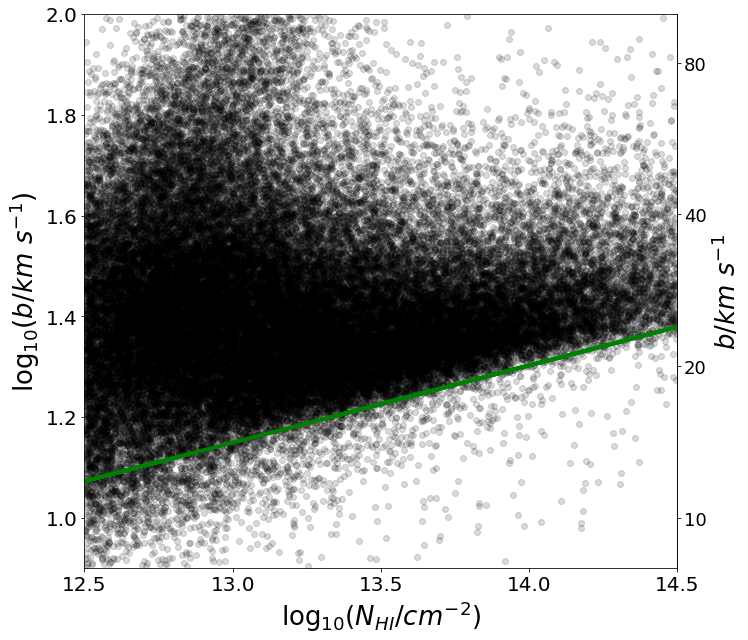}
    \caption{Examples of the distribution of column densities and Doppler widths at $z=2.4$ from the Main simulation with separate transfer functions.
    The green line is the best fit cutoff given by Equation~\ref{eq:cutoff} with $b_0=17.4, \Gamma-1=0.153$.}
    \label{fig:cutoff_fit}
\end{figure}

Perfect spectra and flawless profile fitting are not possible -- Figure~\ref{fig:cutoff_fit} shows a clear trend in the minimum Doppler width with column density, but there are points that lie below the visual cutoff.
To best fit this minimum Doppler width cutoff, \citet{1999MNRAS.310...57S} developed an algorithm which fits the relation
\begin{equation}\label{eq:cutoff}
\log_{10}(b) = \log_{10}(b_0) + (\Gamma-1)\log_{10}(N_{\text{HI}}/N_{\text{HI},0}).
\end{equation}

We follow most closely the algorithm choices used in \citet{2012ApJ...757L..30R}, which includes an initial ($\sigma$) rejection step to remove low points which in an observational setting are most likely metal-contaminated.
Note that in that algorithm the $\sigma-$rejection step uses the RMS deviation while the cutoff fitting step uses the mean absolute deviation and a column density normalization of $N_{\text{HI},0} = 10^{13.6}$, in agreement with \citet{1999MNRAS.310...57S} and \citet{2012ApJ...757L..30R}.
An example of the cutoff fit can be seen in Figure~\ref{fig:cutoff_fit}.

\begin{figure}
    \centering
	\includegraphics[width=\columnwidth]{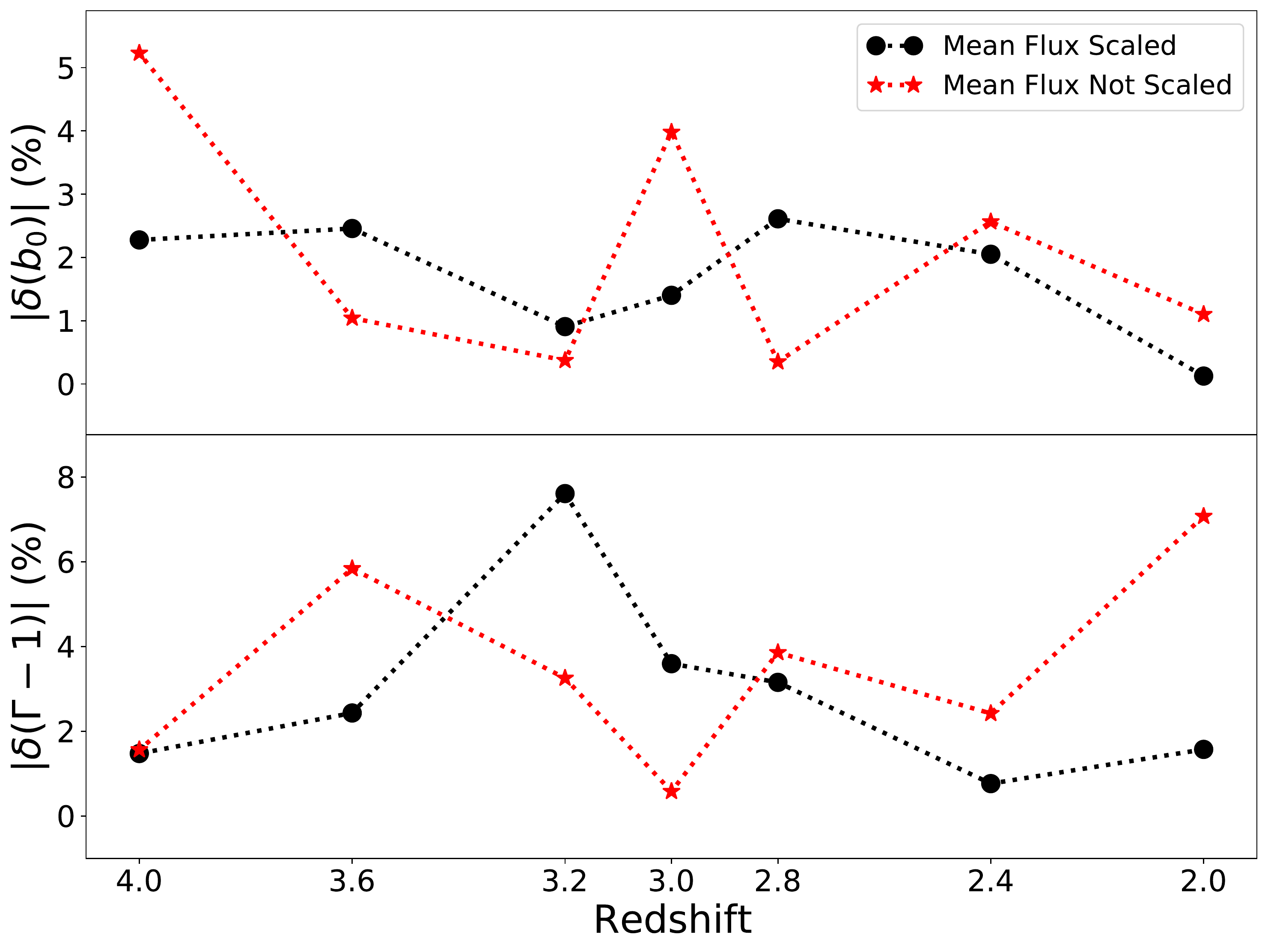}
    \caption{Results for the Doppler width fitted cutoff parameters.
    Shown is the percent absolute difference between the simulations with and without separate transfer functions.
    Top: the difference for the minimum width, i.e. the intercept for the fitted cutoff.
    At most the unscaled results differ by $\approx5\%$ at $z=4$ and the scaled results differ by $\approx2\%$.
    Bottom: same as the top panel, but for the slope of the fitted cutoff.
    The agreement is good, peaking at $\approx8\%$ difference for both the scaled (at $z=3.2$) and unscaled results (at $z=2$).}
    \label{fig:b_results}
\end{figure}

Figure~\ref{fig:b_results} shows the effect separate transfer functions have on the cutoff parameters as a percent difference in the results between the Main simulations.
Difficulties with fitting Voigt profiles at higher redshift due to fewer unsaturated features (a lower ionized fraction) lead to a more ambiguous cutoff in the distribution, so we only attempt to fit a cutoff up to $z=4$.
The difference in the two fit parameters brought about by the separate transfer functions is $2-5\%$ for $b_0$ and $2-8\%$ for $\Gamma-1$. This is similar to the level of convergence with resolution and spectral sample seen in Figure~\ref{fig:b_convergence}, indicating that the primary driver of any differences is most likely due to changes in the sample of spectral features used in the fit. Note that the effect is also not uniform with redshift, leading to an increase in some redshift bins, and a decrease in others.

Our results are consistent with \citet{2012ApJ...757L..30R} to within $5\%$, and \citet{2018ApJ...865...42H} to within $10\%$, at a redshift of $z=2.4$ (where we overlap with both works).
However, the sample used in fitting the lower cutoff has a large effect on the result.
Specifically, the inclusion or omission of a few data points near the cutoff shifts the fit, while points away from the cutoff have little effect.
This is further indicated by the continued variance in the fit parameters as the sample size is increased (see Figure~\ref{fig:nlos_convergence_diff}, bottom two panels).
The variance due to the sample size is on the same order ($<10\%$) as the difference we see between the two Main simulations in Figure~\ref{fig:b_results}.
The effect of the sample used is likely the primary driver of the relatively weak convergence with simulation box size and resolution seen in Figure~\ref{fig:b_convergence}.

Convergence of these results between the Main simulations and the Low Res and Small Box simulations are discussed in Appendix~\ref{appendix:convergence} and can be seen in Figure~\ref{fig:b_convergence}.
Convergence with number of sight lines used is also discussed in the Appendix and can be seen in Figure~\ref{fig:nlos_convergence_diff} (bottom two panels).

\subsubsection{Doppler Width Distribution}\label{sec:bpdf}

Recently, \citet{2019ApJ...876...71H} and \citet{2020arXiv200900016G} have attempted to address the problems associated with fitting the lower cutoff of the $N_{\text{HI}}-b$ distribution.
Specifically, these works avoid fitting the lower cutoff entirely, which removes the strong dependence on just a few points near the cutoff (making poor use of most of the data), and potential systematics in the cutoff fitting algorithm.

\begin{figure*}
    \centering
	\includegraphics[width=2\columnwidth]{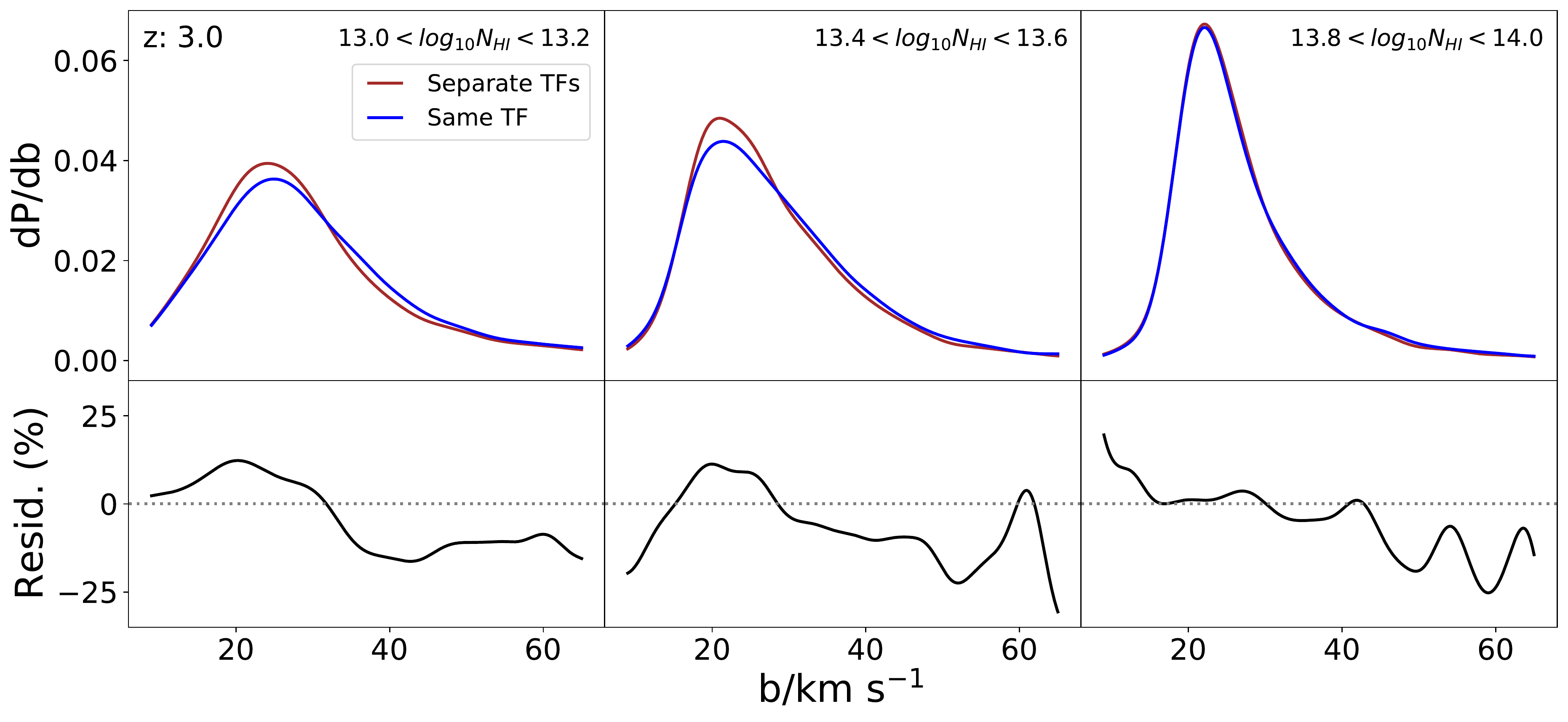}
    \caption{Top: Doppler width distributions at $z=3$ for the three column density bins highlighted in \citet{2020arXiv200900016G}.
    Shown are the distributions for the single transfer function simulation (blue) and species-specific transfer functions (brown) simulation.
    The distributions are from Voigt fitting $5,000$ skewers for each simulation snapshot.
    Bottom: Residuals (as a percent) for the Doppler width distributions between our Main simulations.
    Our simulations differ by less than $1\%$ in mean temperature ($T_0$) and density index ($\gamma$), measured from the $z=3$ snapshot.}
    \label{fig:bpdf}
\end{figure*}

Instead, these methods directly measure the traditional temperature-density parameters (mean temperature $T_0$, and power law index $\gamma$) by comparing the observed $N_{\text{HI}}-b$ distribution to simulations or simulation-derived emulators.
\citet{2019ApJ...876...71H} uses the full two-dimensional $N_{\text{HI}}-b$ distribution and an emulator built on a grid of thermal parameters to estimate the observed $T_0$ and $\gamma$.
\citet{2020arXiv200900016G} bins the Doppler widths by column density, then uses the resulting set of one-dimensional distributions to infer the temperature-density parameters.
Their simulations indicate that the lower column density bins are more sensitive to $T_0$, while the higher column density bins are sensitive to $\gamma$.
The idea is to simultaneously fit the set of observed distributions with simulated distributions (with varying thermal parameters).
An example of the Doppler width distributions is shown in the top panels of Figure~\ref{fig:bpdf}, separated into the three column density bins highlighted in \citet{2020arXiv200900016G}.

To quantify the effect of using separate transfer functions, we find the maximum in the Doppler widths distribution residuals between our Main simulations across the bins shown in Figure~\ref{fig:bpdf}.
From the lower panels in that Figure, we see the (absolute) difference is at most $\approx30\%$.
The difference at other redshifts can be as much as $\approx70\%$ at $z=3.6-4$ ($\approx50\%$ for $z=2-2.4$), down to the $\approx30\%$ seen at and around $z=3$.
The transfer function affects the distribution at a similar level to the residuals \citet{2020arXiv200900016G} found in their fit to observations around $z=3$ (their figure 7).
However, we may not be fully converged with simulation box size or resolution.

The $\approx30\%$ difference brought about by the transfer functions is similar to the difference we find between our Main simulations and the Small Box ($\approx50\%$) and Low Res ($\approx35\%$) simulations (see Figure~\ref{fig:bpdf_convergence}).
It is also similar to the level of convergence \citet{2020arXiv200900016G} found (their figure F1).
Furthermore, the $z=3$ distribution is better converged than the distribution at other redshifts.
The distribution from the Low Res simulations differ from the Main simulations by as much as $80\%$ at lower redshifts ($z=2-2.4$) and as much as $100\%$ at higher redshifts ($z=3.6-4$).
The Box Size simulation distribution also differs by as much as $75\%$ at lower redshifts, and as much as $150\%$ at higher redshifts.
It is not surprising that the higher redshift result is less well converged, since the lower ionized fraction at these redshifts leads to more highly saturated spectra, and thus Voigt fitting is more difficult.
Regardless of these difficulties at higher redshift, the effect of the separate transfer functions on the distribution is on the same order as the effect from simulation box size and resolution at all redshifts explored here.
This indicates that, in terms of the Doppler width distribution, our simulations are not fully converged (see also Appendix~\ref{appendix:convergence}).

\subsection{Flux Power Spectrum}\label{sec:fluxps}

Lyman$-\alpha$ forest spectra from the IGM can also be used to constrain cosmologies alternative to $\Lambda$CDM.
For example, a warm dark matter particle suppresses structure relative to CDM on scales smaller than the WDM particle free-streaming scale \citep{2000ApJ...543L.103N}.
Lyman-$\alpha$ forest spectra probe the scales relevant to the WDM model and can be used to estimate the clumping of matter (the matter power spectrum) through the observed flux distribution (the flux power spectrum) \citep{2004MNRAS.354..684V}.
The flux power spectrum is $P_F(k) = |L^{-1}\tilde{\delta}^2_F(k)|$, where $\tilde{\delta}^2_F(k)$ is the Fourier transform of the flux excess, $\delta_F(k) = F(k)/\langle F(k) \rangle - 1$, and $L$ is the length of the sight lines in velocity space.

The effect of WDM on the flux power spectrum is to suppress high $k$ ($> 0.01$ s/km) power and marginally enhance low $k$ ($< 0.01$ s/km) power \citep{2013PhRvD..88d3502V, 2017PhRvD..96b3522I}.
The shape of the Lyman-$\alpha$ forest flux power spectrum can be used to measure the suppression scale, which directly constrains the WDM particle mass.
The ever increasing number of observed quasar sight lines has meant that a statistically significant sample can be assembled to look at this effect \citep{2017MNRAS.466.4332I, 2018ApJ...852...22W}.

\begin{figure}
    \centering
	\includegraphics[width=0.95\columnwidth]{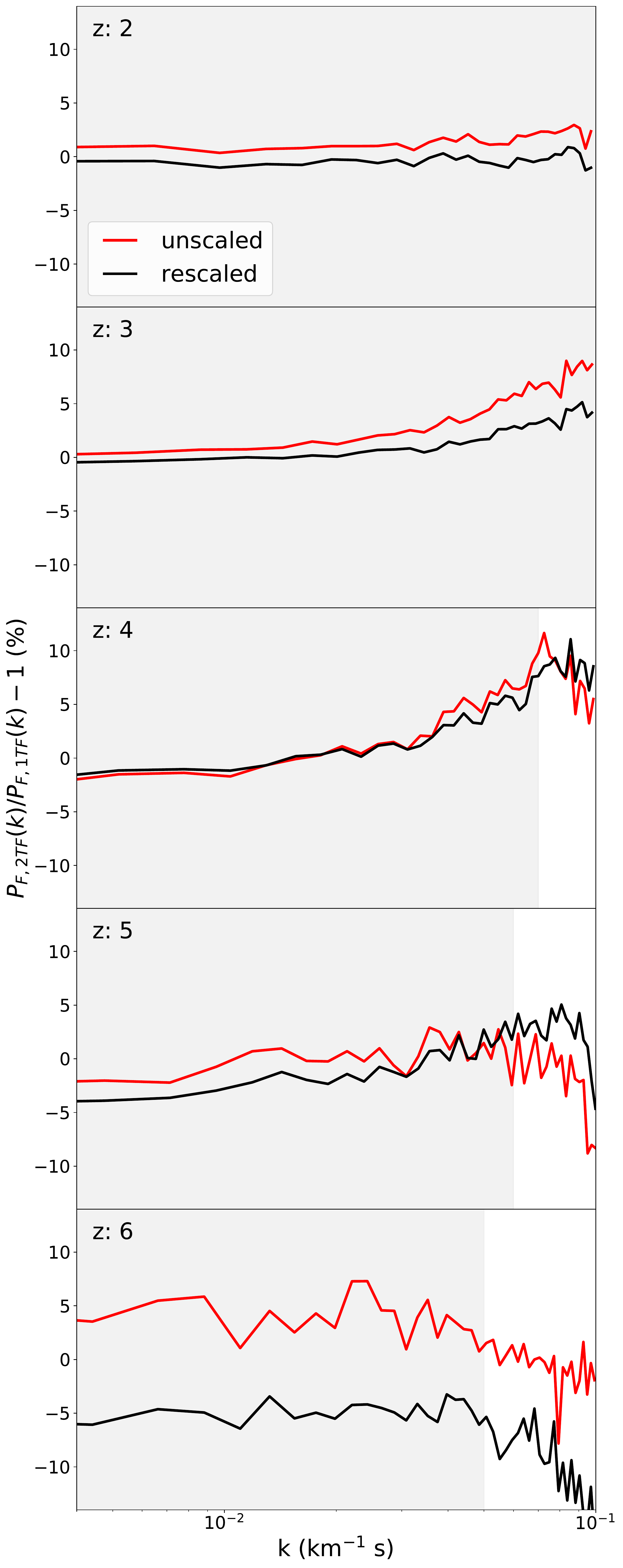}
    \caption{Ratio of flux power spectrum from the simulation with separate transfer functions to the simulation with a single transfer function.
    The red curve shows the result before rescaling the flux, while the black curve is the result after rescaling.
    The grey shaded region in each panel indicates where the flux power spectrum is converged with simulation box size and resolution to at least $10\%$ (see Figure~\ref{fig:fps_convergence_scaled}).
    On the scales probed here, the effect at $z>2$ tops out at $\approx5\%$, while at $z=2$ the effect is $<1\%$.}
    \label{fig:fps_results}
\end{figure}

These constraints rely on accurate modeling of the flux power spectrum in simulations using a CDM or WDM model, coupled with observed Lyman-$\alpha$ forest spectra.
Figure~\ref{fig:fps_results} shows the effect of using separate transfer functions for baryons and CDM on the Lyman-$\alpha$ forest flux power spectrum.
Shown is the percent change in the power spectrum when using separate transfer functions instead of a single transfer function.
The effect is generally strongest at the high end of the $k$ range, with a decrease in power across all redshifts at $k<0.02$ for the rescaled result.
The effect increases with redshift, however both the rescaled and unscaled results remain at $\lesssim5\%$ across the range of $k$ our simulations reliably probe.

\begin{figure}
    \centering
	\includegraphics[width=\columnwidth]{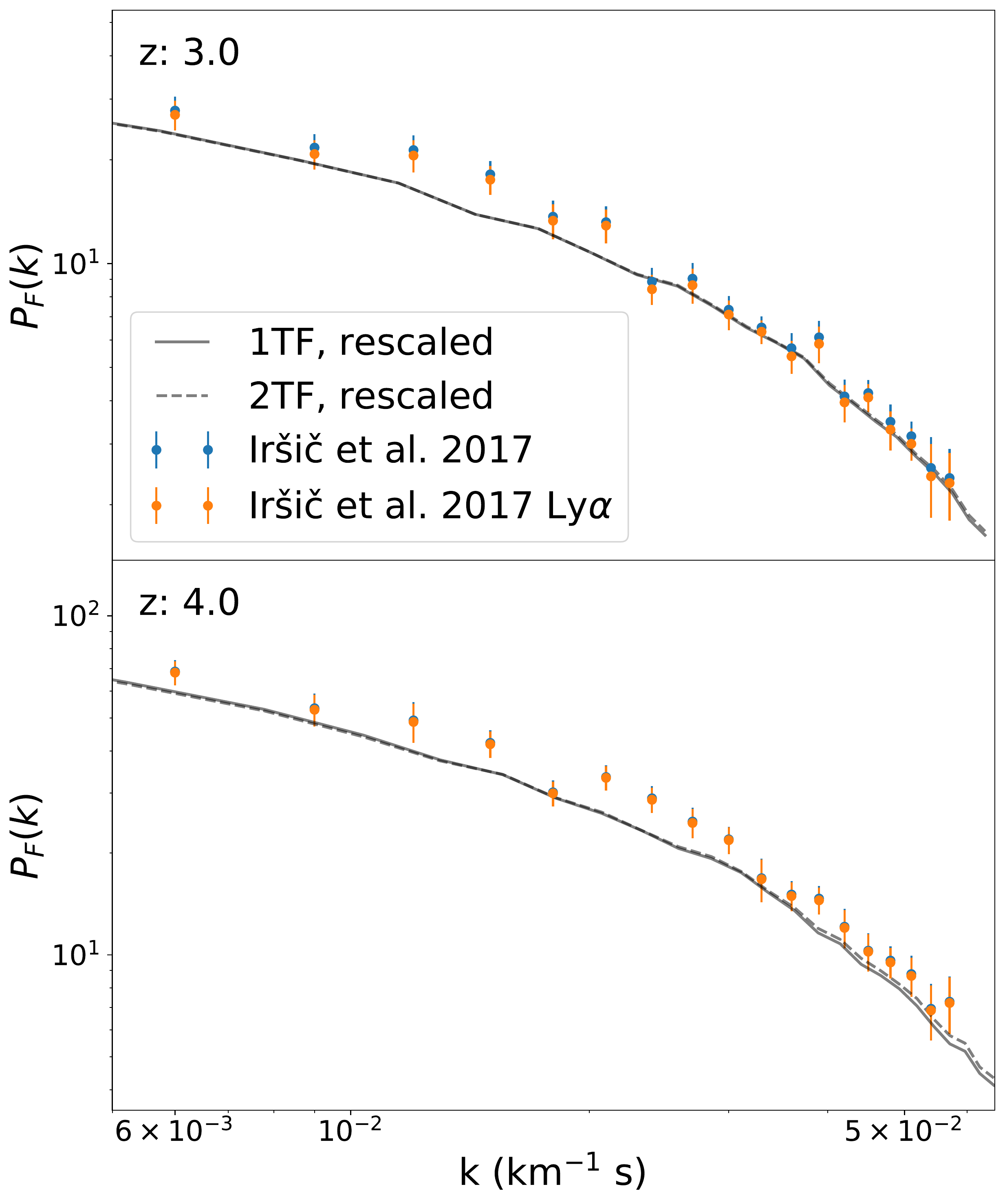}
    \caption{Comparison of the flux power spectra from the simulations presented here, to the data presented in \citet{2017MNRAS.466.4332I}.
    In \citet{2017MNRAS.466.4332I} they present the full flux, as well as an estimate on the contribution from only Lyman-$\alpha$, which is comparable to our simulated results.
    For low $k$, the simulations appear to underpredict the power with the largest discrepancies at $\approx2.5\sigma$, but most lie within $1\sigma$.}
    \label{fig:irsic_results}
\end{figure}

While observations extending to the largest wave numbers used here are not presently available, we can compare the flux power spectrum we obtain from our simulations to currently available data where they overlap.
Figure~\ref{fig:irsic_results} shows our flux power spectrum at $z=3$ and $z=4$ compared to data taken from \citet{2017MNRAS.466.4332I}.
Their estimate of the Lyman-$\alpha$ contribution to their total flux power, as well as their total flux power, are shown with the reported errors.
Our results show a small underestimate of the power in comparison with them, but are roughly consistent.
To estimate the level of agreement, we interpolate our flux power spectrum onto the wave numbers of the \citet{2017MNRAS.466.4332I} data.
We find that at most, our power spectrum differs by $\sim2.5\sigma$ from their data (where the deviation is their reported statistical and systematic errors, added in quadrature).
For most data points, the difference is within $1\sigma$.
In terms of this deviation, the difference between the single and separate transfer function cases is $<1\sigma$.

\subsection{Matter Power Spectrum}\label{sec:matterps}

The total matter power spectrum is affected only at the $1-2\%$ level at all scales probed with the simulations presented in this work ($k=0.6-100$ h Mpc$^{-1}$).
This is unsurprising, as the effect of the separate initial conditions is to reproduce the offset of the power between the baryons and dark matter, and not to change the total matter power spectrum.

\begin{figure}
    \centering
	\includegraphics[width=\columnwidth]{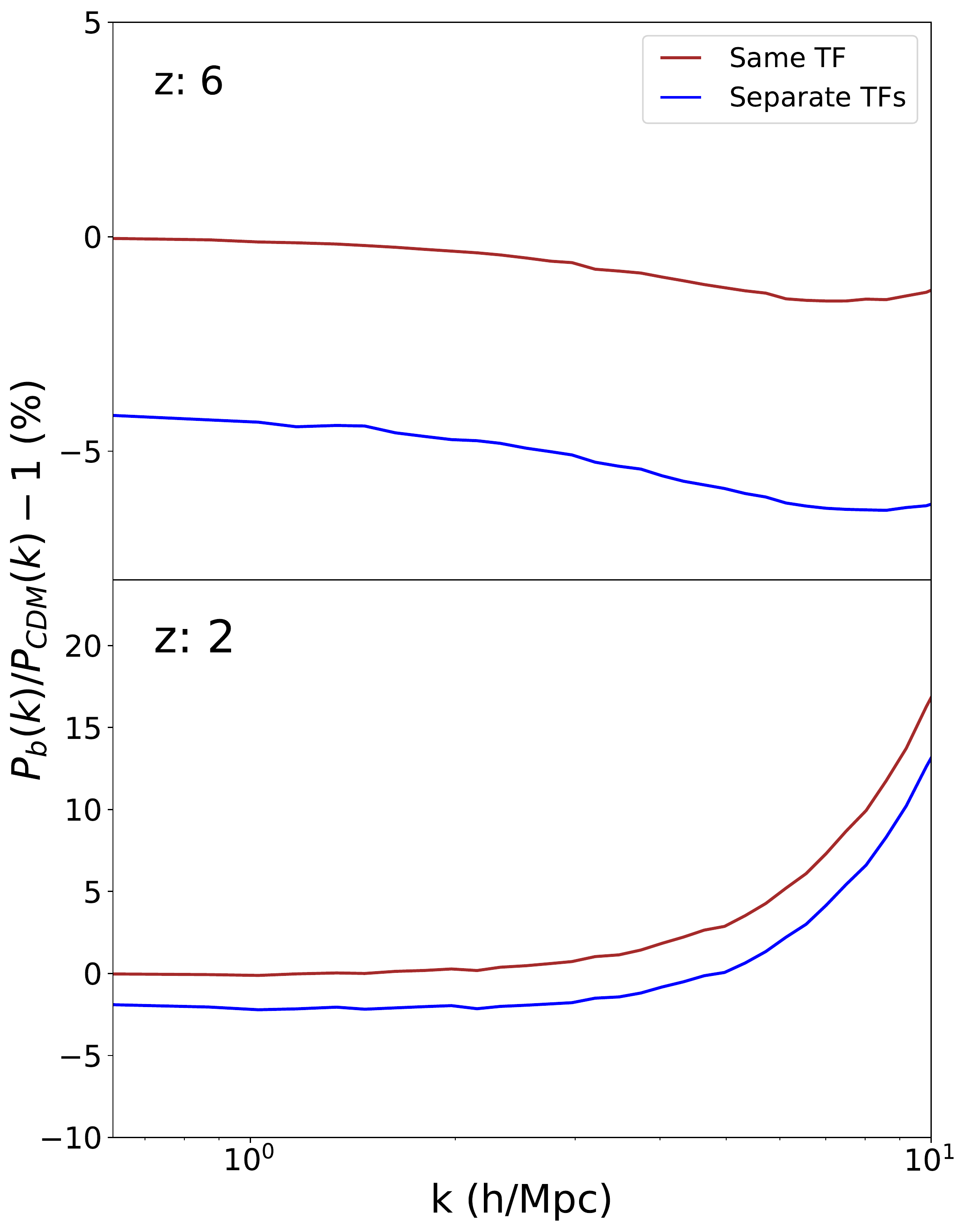}
    \caption{Percent difference in the ratio of the species specific matter power spectrum for baryons over CDM.
    At $z=2$ and $z=6$ the simulations using a single transfer function predict more power in the baryons than the case with separate transfer functions.
    At $z=6$, in the separate transfer function simulation we see higher power in the CDM, as expected  (meanwhile the single transfer function simulation is consistent with $P_b \approx P_{CDM}$).
    At $z=2$ we see the expected enhancement in baryon power on small scales.}
    \label{fig:mps_species}
\end{figure}

Figure~\ref{fig:mps_species} shows the difference in the species specific matter power spectrum ratio (baryon power over CDM power).
From this we see that the effect of the separate initial conditions is to decrease the power in the baryons, while retaining the behaviour at both higher redshift (for linear structure to dominate on a large range of scales) and at lower redshift (for baryons to collapse into non-linear structures at small scales).
The offset is independent of $k$, and is consistent both with linear theory \citep{2012ApJ...760....4O} and with \citet{2020arXiv200200015B}.

We ran two additional simulations, both with the same volume and particle number as the Small Box simulation.
The first used species-specific transfer functions, but initialized both the baryons and CDM using (offset) grids.
Previous work \citep{2020arXiv200200015B} has shown that without additional adjustments \citep{2020arXiv200809124H, 2020arXiv200809123R}, this setup can lead to artificial growth in the CDM-baryon power difference.
On the scales probed in this work, we find that the simulation using offset grids does a fairly good job in reproducing the power difference, agreeing with our separate transfer function Main simulation to within $2\%$ on all the but the smallest scales ($k > 7$ h Mpc$^{-1}$, where it stays within $5\%$).
This indicates that the artificial growth in the power difference is sourced from larger scale velocity perturbations, and that the half-glass approach is unnecessary on the scales probed here.

The second additional simulation included WDM (a $5.3$ keV WDM particle, consistent with \citet{2017PhRvD..96b3522I, 2020JCAP...04..038P}).
The change in the total matter power spectrum remained at the sub-percent level for all scales.
The baryon-DM power difference agreed with the CDM simulations up to $k \approx 3-4$ h Mpc$^{-1}$, with a relative drop in power ($\sim3\%$) at higher $k$, consistent with expectations (WDM suppressing small scale structure).
This drop-off was consistent between the simulations with and without species specific transfer functions, indicating that the effect of using separate transfer function is independent of the DM temperature.

\section{Summary \& Conclusions}\label{sec:conclusions}

In this work we have explored how switching from a single initial transfer function (for both baryons and CDM) to species specific transfer functions affects properties of the IGM.
Using a set of high resolution simulations, we have quantified the effect this change has on probes of the IGM via the Lyman-$\alpha$ forest.
The Main simulations presented here differ only in the transfer functions used; one uses a single transfer function, the other follows \citet{2020arXiv200200015B}, adopting species specific transfer functions.
Our work is motivated by simulations sometimes failing to match the theoretical offset between baryon and cold dark matter power, though we found that on the scales probed in this work ($k=0.6-10$ h Mpc$^{-1}$), standard simulation methods did fairly well.
Artificial spectra were extracted from snapshots in the $2<z<6$ range and statistics relevant to the thermal history of the IGM and WDM models were calculated.
Below we summarize the results of this work.

\begin{itemize}
    \item The curvature statistic is relatively unaffected by the use of the species specific initial transfer functions, with a peak difference of $<2\%$ at $z=6$.
    The effect of the transfer functions (on average, shifting log$_{10}\eta$ by $\approx0.01$) is about the same as the $1\sigma$ ($2\sigma$) uncertainty observed in \citet{boera14} (\citet{2011MNRAS.410.1096B})
    \item The Doppler width cutoff fit parameters continue to vary with number of sight lines used, even when using a large number of sight lines (see Figure~\ref{fig:nlos_convergence_diff}).
    This is likely due to the fitting method, which depends strongly on the few data points near the cutoff.
    The effect on these parameters is at most $\approx5\%$ for the fit intercept and $\approx8\%$ for the fit slope, however the variance due to sample size has a similar level of effect.
    The effect of the transfer functions seen here is at a similar level to the error \citet{2012ApJ...757L..30R}, and generally less than the error \citet{2018ApJ...865...42H} found from fitting to observations.
    \item The Doppler width distribution is affected by $\approx30\%$ around $z=3$, which is similar to the difference between observation and best fit in \citet{2020arXiv200900016G}.
    However our simulations at $z=3$ are only converged to $\approx50\%$ with box size and $\approx35\%$ with resolution, the same order as the effect from the transfer functions.
    \item The flux power spectrum is affected more at high $k$ ($k > 0.05$ s/km) and redshift ($z>2$).
    However, the enhancement to the power is at most $\approx5\%$ for $z>2$ and $\lesssim1\%$ at $z=2$.
    This level of effect is small compared to observational uncertainties, amounting to $\approx10\%$ of the uncertainty presented in \citet{2017MNRAS.466.4332I} (and see Figure~\ref{fig:irsic_results}).
\end{itemize}

For measures of the thermal state of the IGM, the effect of separate transfer functions is either small ($\sim1\%$ for the curvature), or subdominant to either variance inherent to the method of calculation (Doppler width cutoff) or convergence (Doppler width distribution).

The flux power spectrum is relatively unaffected on the scales and times which are currently well observed, however our results indicate that it may become important on smaller scales or higher redshift.
This may indicate that using separate transfer functions may be important for future observations and surveys.
However, the effect is most pronounced at early times and on small scales, which constrain WDM most effectively.
The importance of this effect will only increase as future measurements lead to a higher resolution flux power spectrum, constraining WDM models more stringently.

The future study of the IGM will be predicated on measuring absorption spectra at higher redshifts and at higher resolution using large optical and infrared telescopes in conjunction with broader surveys such as the James Webb Space Telescope \citep{2019BAAS...51c.440B}.
Given the ever increasing sample size and quality of IGM observations, it is paramount that simulations keep pace by improving their precision and modeling.
The adjustment to the initial conditions in simulations explored here is one such improvement, but there are others which should be implemented as well, for example the modeling of He~{\sc ii} reionization \citep{2020MNRAS.496.4372U}.
Improved simulations, in concert with future observations, will push the study of the IGM into the reionization epoch as it occurs, leading to a greater understanding of this relatively recent major phase transition, as well as the formation of the first galaxies and their subsequent evolution.

\section*{Acknowledgements}

This material is based upon work supported by the National Science Foundation Graduate Research Fellowship under Grant No. DGE-1326120.
SB and PUS were supported by NSF grant AST-1817256.

Computing resources were provided by NSF XSEDE allocation AST200018.
The authors acknowledge the Frontera computing project at the Texas Advanced Computing Center (TACC) for providing HPC and storage resources that have contributed to the research results reported within this paper.
Frontera is made possible by National Science Foundation award OAC-1818253.
URL: \url{http://www.tacc.utexas.edu}

\section*{Data Availability}
Flux and matter power spectra for select redshifts, as well as MP-Gadget parameter files for each simulation, are available at \url{https://github.com/mafern/TF_data}. Artificial spectra from the simulations are available upon request.
The simulation data underlying this article can be reproduced using the public code at \url{https://github.com/MP-Gadget/MP-Gadget}.

\bibliographystyle{mnras}
\bibliography{refs.bib}

\appendix

\section{Convergence}\label{appendix:convergence}

We check the convergence of our simulations with box size and mass resolution by running an additional four simulations.
For both the separate and same transfer function cases we run a lower mass resolution simulation (Low Res) and a smaller box size simulation (Small Box).
The simulation volume, particle number, and mass resolution can be seen in Table~\ref{table:simulations}.
The mass resolution used in the Main simulations agrees with \citet{2011MNRAS.410.1096B}, which previously showed convergence for the curvature at that resolution.
\citet{2014MNRAS.438.2499B} showed convergence for the $N_{\text{HI}}-b$ cutoff parameters using the same set of simulations.

\begin{figure}
    \centering
	\includegraphics[width=\columnwidth]{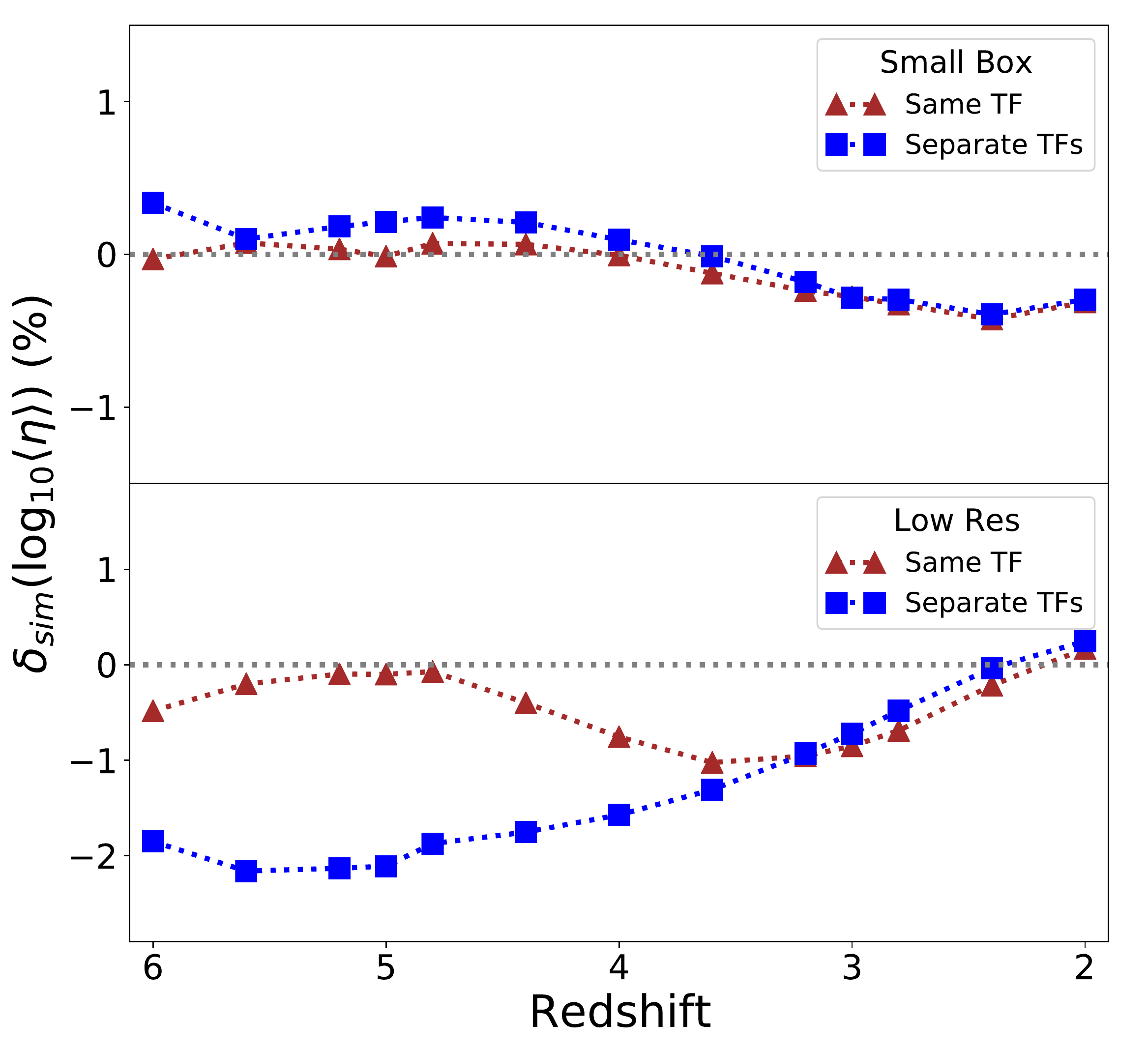}
    \caption{Convergence for the curvature.
    The convergence is quite good, staying within $\sim1\%$ for the smaller box simulations, and within $\sim2\%$ for the lower resolution simulations.
    This emphasizes the importance of the mass resolution in simulations aimed at probing the Lyman-$\alpha$ forest.}
    \label{fig:curvature_convergence}
\end{figure}

Figure~\ref{fig:curvature_convergence} shows the fractional difference as a percentage between each of the Main simulations, and the two associated convergence simulations (called $\delta$) for the curvature.
The curvature calculated here uses spectra without added noise, obviating the need for a spline fit.
Otherwise, the calculation is the same as that outlined in Section~\ref{sec:curvature} (spectra are renormalized into $10$ Mpc/h sections and the mean flux is rescaled).
The curvature is well converged, with a difference of $\lesssim1\%$ for the Small Box simulations, and $\sim2\%$ for the Low Res simulations.

\begin{figure}
    \centering
	\includegraphics[width=\columnwidth]{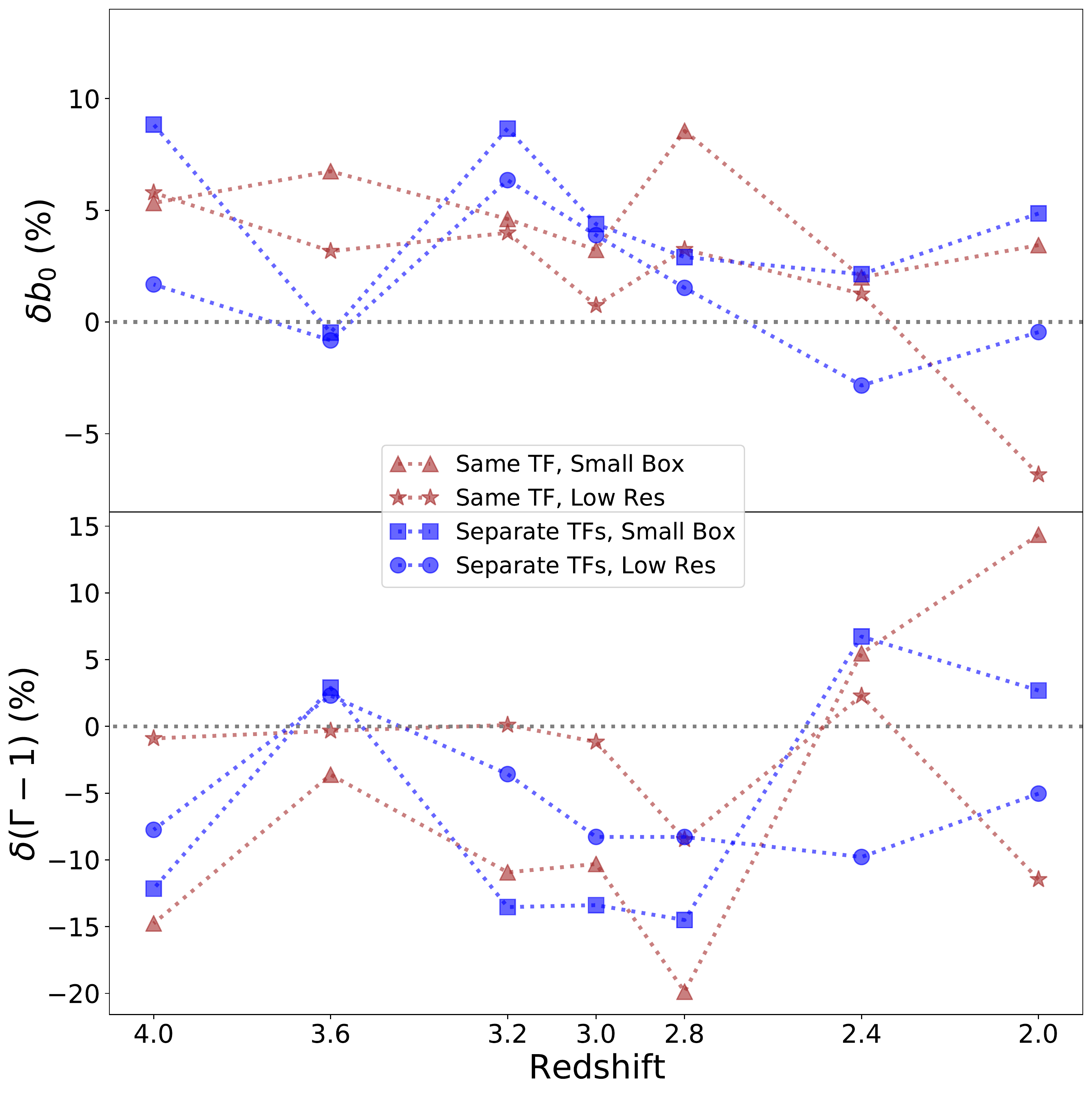}
    \caption{Convergence for the Doppler width cutoff fit values.
    Top: the percent difference for the minimum width (cutoff intercept) fit values between the Main simulations and the Low Res and Small Box simulations.
    The convergence stays within $10\%$.
    Bottom: same as the top panel, but for the slope of the cutoff.
    The convergence is less good, staying within $20\%$.}
    \label{fig:b_convergence}
\end{figure}

Figure~\ref{fig:b_convergence} shows the convergence for the two fit parameters of the $N_{\text{HI}}-b$ cutoff.
As this method uses a population of features taken from each set of spectra, renormalizing between different volume simulations is not necessary.
The logarithmic intercept, $b_0$, has converged to an absolute difference of $<10\%$ at all redshifts.
The logarithmic slope, $\Gamma-1$, is less well converged, remaining within an absolute difference of $< 20\%$ at all redshifts.
However, the amount of data used to make the fit strongly affects the result.
This can be seen in the bottom two panels of Figure~\ref{fig:nlos_convergence_diff}, where the convergence with the number of sight lines used is shown.
In contrast with the flux power spectrum and curvature, these fit parameters do not completely converge, instead exhibiting some variance all the way up to the inclusion of all $5,000$ sight lines.

\begin{figure*}
    \centering
	\includegraphics[width=2\columnwidth]{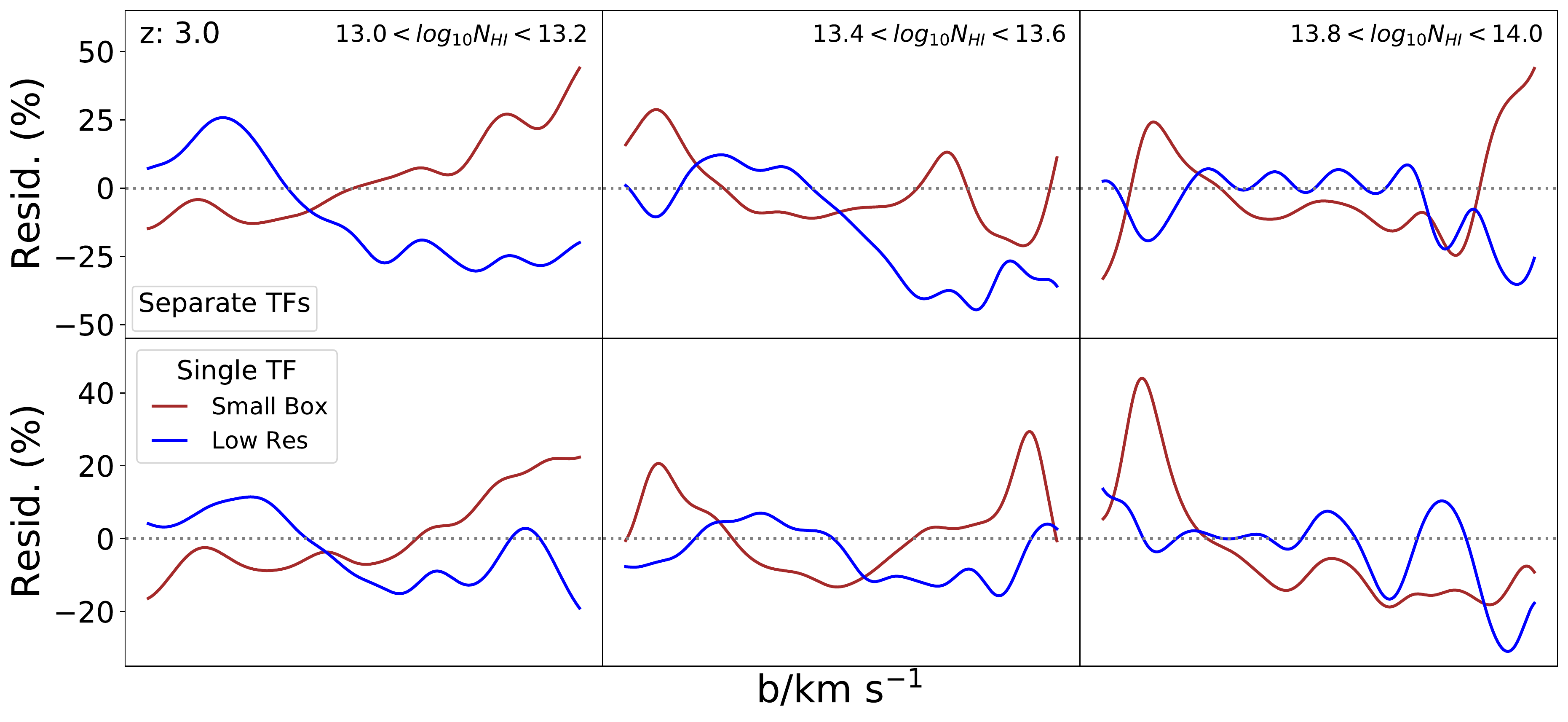}
    \caption{Convergence of the Doppler width distribution at a redshift of $z=3$. 
    Top: the percent difference between our Main simulation and the Small Box and Low Res simulations, for the simulations using species specific transfer function.
    Bottom: same as the top panel, but for the simulations using a single transfer function.}
    \label{fig:bpdf_convergence}
\end{figure*}

Figure~\ref{fig:bpdf_convergence} shows the convergence at $z=3$ for the Doppler width distribution.
Specifically, Figure~\ref{fig:bpdf_convergence} shows the percent difference between the Doppler width distributions for the Main and Small Box simulations, and between the Main and Low Res simulations.
The top panels show this difference for simulations using species specific transfer functions, while the lower panels show this for the simulations using a single transfer function.
The maximum difference across all three bins associated with the Small Box simulation is $\approx50\%$, while the maximum difference associated with the Low Res simulations is $\approx35\%$.
This is a similar level of convergence to \citet{2020arXiv200900016G} (their Figure F1).
However, as stated in Section~\ref{sec:bpdf}, the level of convergence shown here is on the same order as the effect seen from using separate transfer functions ($\sim25\%$).
We also note that redshifts around $z=3$ are better converged, with the Low Res and Small Box simulations differing by as much as $\approx80\%$ for redshifts between $z=2$ and $z=2.4$, and at higher redshifts ($z=3.6-4$) they differ by as much as $\approx150\%$.
This indicates that for the Doppler width distribution, we are not fully converged with resolution or box size.

\begin{figure}
    \centering
	\includegraphics[width=\columnwidth]{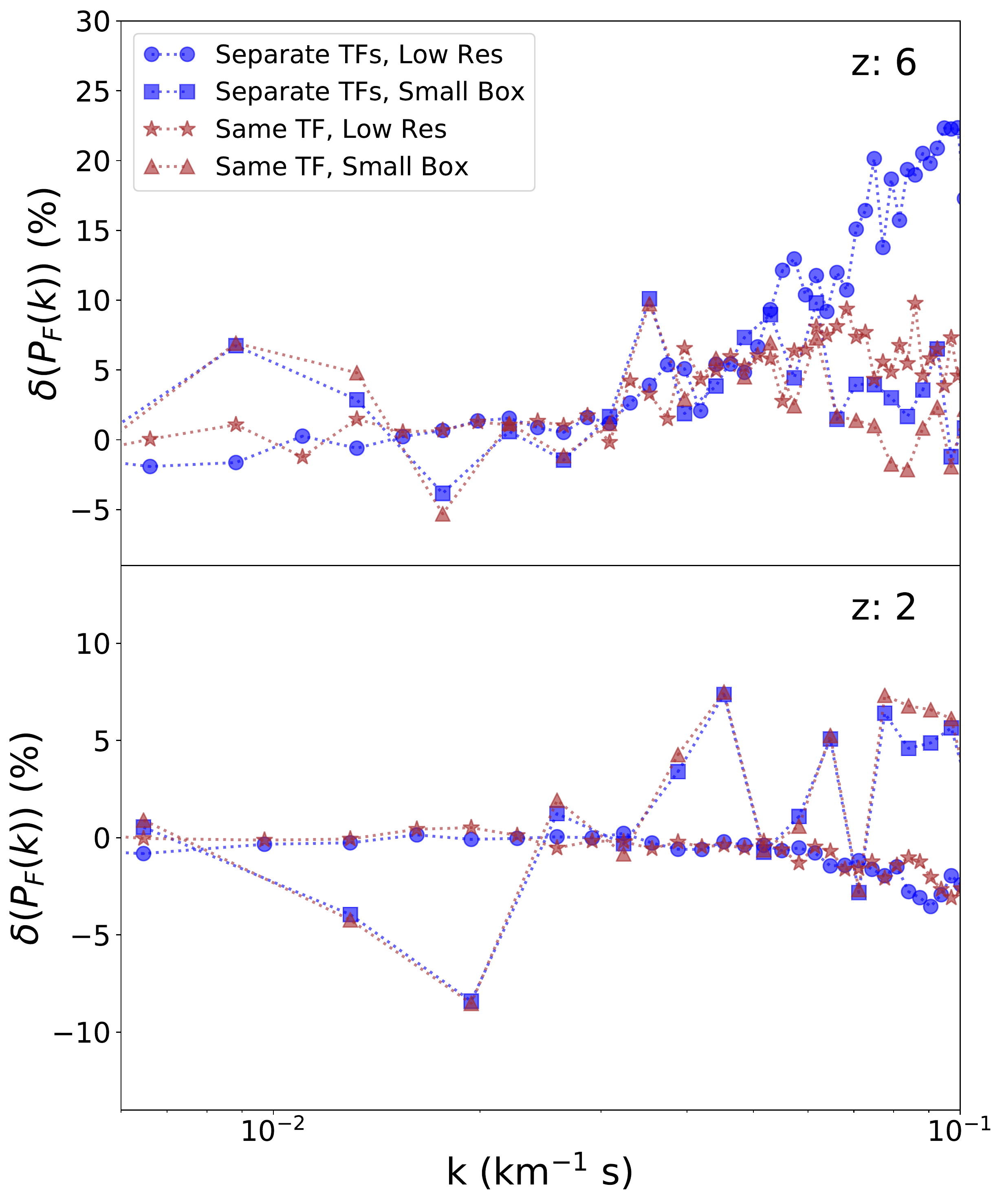}
    \caption{Convergence of the flux power spectrum with simulation box volume and gas mass resolution.
    At early times ($z=6$) we are converged within $10\%$ up to $k\approx0.05$ km$^{-1}$s.
    For larger $k$, the Low Res, separate transfer function simulation is less well converged.
    At late times ($z=2$) we are fully converged with resolution, however the convergence with box size is less good ($\lesssim10\%$).}
    \label{fig:fps_convergence_scaled}
\end{figure}

\begin{figure}
    \centering
	\includegraphics[width=\columnwidth]{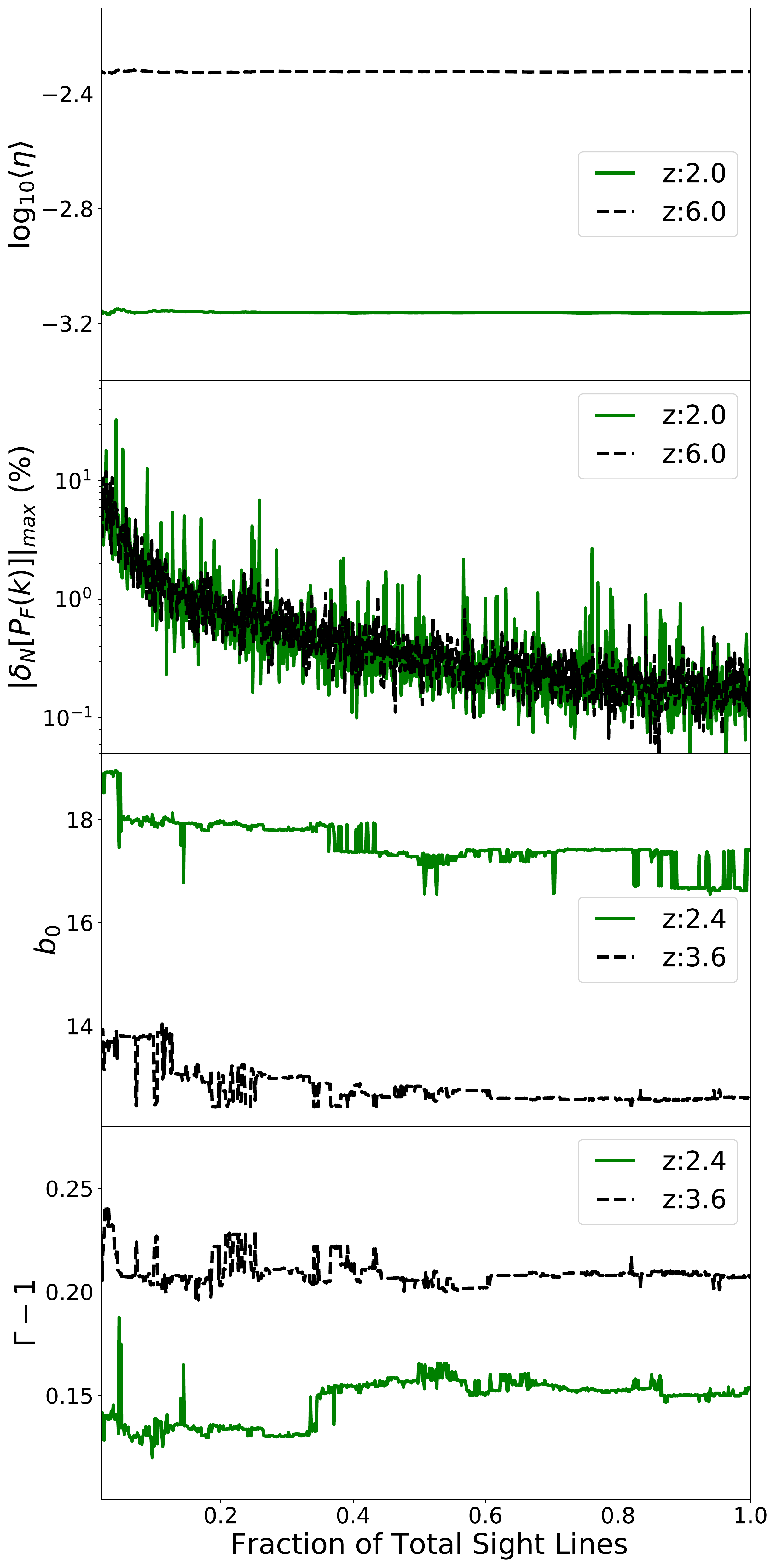}
    \caption{Convergence of curvature, flux power spectrum, and $N_{\text{HI}}-b$ cutoff parameters with number of artificial spectra used.
    Shown here is the convergence for simulations with separate transfer functions.
    The flux power spectrum convergence shows the maximum change in the spectrum when additional sight lines are added.
    Note that the sight lines are randomly placed in the simulation box.}
    \label{fig:nlos_convergence_diff}
\end{figure}

Figure~\ref{fig:fps_convergence_scaled} shows the convergence for the flux power spectrum at the two redshifts which span our analysis.
We are well converged at low $k$ ($<0.03$ km$^{-1}$s) for all simulations and redshifts.
At $z=6$ we are well converged ($<10\%$) for three cases, while the lower resolution simulation with separate transfer functions is not well converged beyond $k\approx0.05$ km$^{-1}$s.
The shaded regions in Figure~\ref{fig:fps_results} which indicate the trusted $k$ range for each redshift are based on this.
At $z=2$ we are well converged across the range ($k<0.1$ km$^{-1}$s) explored here.

The matter power spectrum is converged with gas mass resolution at all redshifts ($2 < z < 6$) and scales ($k=0.6-100$ h Mpc$^{-1}$) such that the difference between the higher and lower resolution simulations is $\lesssim5\%$.
Convergence with box volume is $\lesssim20\%$ in the range $k=10-100$ h Mpc$^{-1}$, and $\approx30\%$ from $k=0.6-10$ h Mpc$^{-1}$.

Finally, we check the convergence of the curvature, flux power spectrum, and $N_{\text{HI}}-b$ cutoff fit parameters with the number of sight lines used in each of their calculation.
Figure~\ref{fig:nlos_convergence_diff} shows this convergence for the Main simulation using separate transfer functions.
The convergence trends for all quantities are also seen in the simulations using a single transfer function.
Each statistic is calculated using only the corresponding fraction of the $5,000$ sight lines available (e.g. the $0.2$ value for the $N_{\text{HI}}-b$ cutoff fit parameters use only features from the first $1000$ random sight lines).
While the curvature is insensitive to the number of sight lines used (beyond $\sim 50$ sight lines), the flux power spectrum depends strongly on the number.
The convergence of the flux power spectrum shows the maximum difference between consecutive power spectra (the difference between the spectrum with $x$ sight lines and the spectrum with $x+5$ sight lines).
The $N_{\text{HI}}-b$ cutoff fit parameters continue to fluctuate even with a large number of sight lines.
The variance, even when using all of the sight lines in the $N_{\text{HI}}-b$ cutoff fit parameters may explain the worse convergence with simulation size and resolution seen in Figure~\ref{fig:b_convergence}.

\bsp
\label{lastpage}
\end{document}